\documentclass[journal]{vgtc}                     

\onlineid{0}

\vgtccategory{Research}

\vgtcpapertype{please specify}

\title{RekomGNN: Visualizing, Contextualizing and Evaluating Graph Neural Networks Recommendations}

\author{%
  Camelia D. Brumar,
  Gabriel Appleby, 
  Jen Rogers,
  Teddy Matinde,
  Lara Thompson,
  Remco Chang,
  Anamaria Crisan
}

\authorfooter{
  \item {
        Camelia D. Brumar, Gabriel Appleby, Jen Rogers, and Remco Chang are with Tufts University.
        E-mail: \{camelia\_daniela.brumar, gabriel.appleby\}@tufts.edu, \{jen, remco\}@cs.tufts.edu, 
        }
  \item {
        Teddy Matinde, Lara Thompson, and Anamaria Crisan are with Tableau.
        E-mail: \{tmatinde, lara.thompson, ana.crisan\}@salesforce.com
  }
}

\newcommand{\sys}{RekomGNN\xspace}

\abstract{%
  Content recommendation tasks increasingly use Graph Neural Networks, but it remains challenging for machine learning experts to assess the quality of their outputs. Visualization systems for GNNs that could support this interrogation are few. Moreover, those that do exist focus primarily on exposing GNN architectures for tuning and prediction tasks and do not address the challenges of recommendation tasks. We developed \sys, a visual analytics system that supports ML experts in exploring GNN recommendations across several dimensions and making annotations about their quality. \sys straddles the design space between Neural Network and recommender system visualization to arrive at a set of encoding and interaction choices for recommendation tasks. We found that \sys helps experts make qualitative assessments of the GNN's results, which they can use for model refinement. Overall, our contributions and findings add to the growing understanding of visualizing GNNs for increasingly complex tasks.
}


\teaser{
  \centering
  \includegraphics[width=\linewidth]{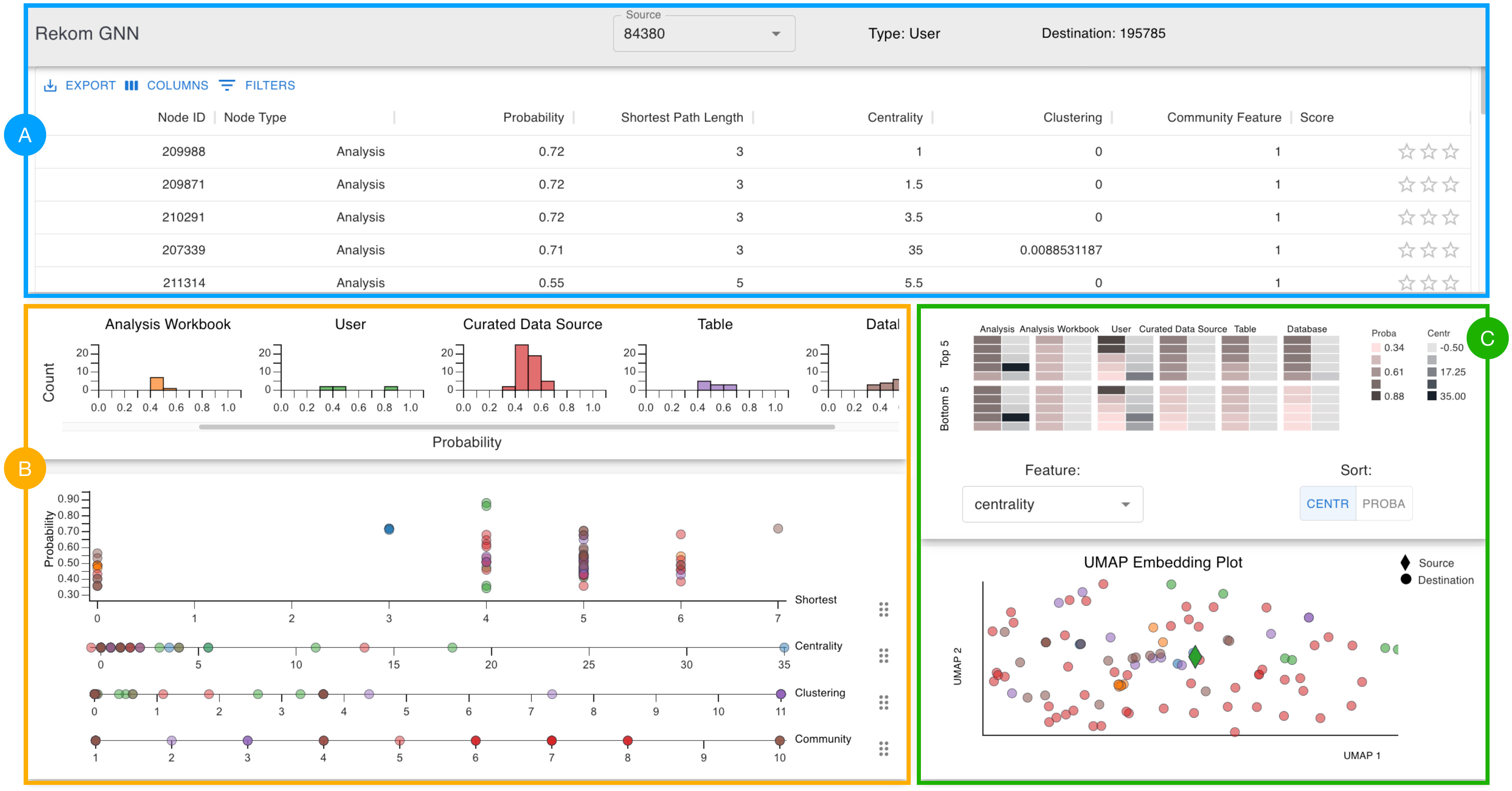}
  \caption{%
  	The RekomGNN interface compromises three components. (A) A Data Panel that allows users to examine a plain view of recommended nodes and highlight recommendations of interest across the system. Users can also score and later export the quality of the recommendations within the data panel allowing for further external refinement of the GNN. (B) A Recommendation Overview Panel compares prediction probabilities across different node types and different attributes of individual nodes and the input graph. (C) A Recommendation Detail Panel that allows users to further contextualize the relationship between recommendations, prediction probabilities, and different features in relation to other nodes.%
  }
  \label{fig:teaser}
}

\graphicspath{{figs/}{figures/}{pictures/}{images/}{./}} 

\usepackage{tabu}                      
\usepackage{booktabs}                  
\usepackage{lipsum}                    
\usepackage{mwe}                       

\usepackage{mathptmx}                  

\begin{document}



\maketitle

\section{Introduction}\label{intro}

Modern data analysis requires practitioners to navigate increasingly complex repositories to procure data~\cite{Terrizzano:CODAR:2015DataWT,Fernandez:aurum:2018}. While original big data repositories primarily comprised relational datasets, modern alternatives, such as data lakes and attendant catalogs, offer the ability to store a greater variety of content with more flexible query and retrieval options~\cite{hai:datalakedefine:2021}. However, the combination of both more and varied data introduces significant overhead for analysts who make use of these resources. For example, an analyst's data discovery attempts typically require a time-consuming iterative trial-and-error process~\cite{Crisan:recon:2019}. They may also seek more than just a dataset to initiate their analysis, for example, they may be interested in related analysis attempts to inform and inspire them~\cite{Raghunandan:LodestarSI:2022}. Developing a viable solution to support analysts situating themselves in such large and diverse repositories of analytic content is complex and requires a multifaceted approach.

In this research, we explore the challenges of developing content recommendation systems for such modern repositories. Content recommendation approaches have been widely used in commercial applications~\cite{Lops:ContentRec:2011,Schafer:CFRec:2007,Burke:HybridRec:2002}, but their utility to data analysis workflows have been under-explored. Recommendation approaches that do exist for analytic workflows are tailored more to the downstream analyses process, such as preparation~\cite{Cong:AutoSuggest:2020}, exploration~\cite{Lee:LUX:2021}, and visual analysis~\cite{kam:voyager:2016} with few solutions tackling the content discovery problem. To improve content \textit{search} in large repositories, prior research has proposed leveraging knowledge graphs\cite{Fernandez:aurum:2018} and catalog type metadata~\cite{Halvey:GOODS:2016} -- but these approaches still impose a large overhead to analysts since they must still construct useful queries. To approach the problem of \textit{recommending analytic content}, we collaborated with software and machine learning engineers (SEs and MLEs, respectively) that were exploring the potential utility of Graph Neural Networks (GNNs).

The application of GNNs to recommendation tasks is a fairly recent but promising branch of research~\cite{Wu:GNNRec_Overview:2022,Lops:ContentRec:2011,KonstanTerveen:HumanCenteredRecommenderSystems:2021}. The use of GNNs is appealing because it can leverage already existing graph structures,  like knowledge graphs or the provenance structure of catalogs, to recommend analytic content. However, it is also challenging to develop an appropriate GNN solution and vet the quality of its results. Together with our collaborators, we developed~\sys, a visual analysis tool for interrogating GNN results in the context of analytic content recommendation. We examine the trade-offs in the broad and joint design space of GNN and recommendation system visualization to justify the design choices of~\sys. \textbf{We present this collaboration as a design study that makes the following contributions:}

\begin{itemize}
    \item A data and task abstraction for the application of GNNs to content recommendation for data analysis
    \item The design and development of \sys, which supports the interrogation of GNN results for recommendation tasks
    \item Usage scenarios and an evaluation of \sys with machine learning and software engineers that examines design trade-offs
\end{itemize}

Our research and its findings offer generalizable insights into the challenges of recommending analytic content by leveraging emergent machine learning techniques and modern data repositories.

\vspace{.5em}
\section{Domain Background and Collaboration Context}\label{bg_and_domain}
In this section, we elaborate on our collaborators' challenges in the context of Graph Neural Networks (GNNs) and Recommendation tasks. We provide a high-level overview of GNNs to convey both the pain points of our collaborators and information to visualize.

\subsection{Defining Analytic Content}\label{analytic_content}
Relational databases housed within data warehouses represent the majority mechanism for managing so-called ``big data'', but more modern advances enable the storage of a greater variety of content~\cite{Strohbach:warehouse:2016}. For example, not only are datasets stored, but so are their downstream analyses. These analyses can exist in the form of computational notebooks, dashboards, and other types of documents. The uses of these data in computational pipelines that then feed into downstream data analyses can also be stored. The relationships between these different content types, from data, pipeline workflows, and analyses are captured via a lineage graph. In Section~\ref{data_overview} and Figure~\ref{fig:input_grapj} we provide a specific example of such a linage graph from our collaborators'.

In addition to the lineage graph, repositories also store metadata about the content, for example, what it is, when it was created, who it is owned by, and some additional properties; for a tabular dataset, these additional metadata properties can include information about specific attributes (name, domain, possible user added tags). In some instances additional data may also be derived by analyzing the content itself, for example, to identify which attributes were used in analyses. When repositories are instrumented for usage telemetry it is also possible to ascertain the popularity of content. The collective information about the different types of content and their relationships to one another can be leveraged to help identify pertinent content to an analyst. However, doing so remains an open problem.

\subsubsection{Motivating Scenario}
The following scenarios are intended to concretely convey the challenges faced by both analysts and our collaborators when managing large amounts of analytic content. First, we describe the situation faced by a potential  end-user needing to find content for data analysis:
\begin{quote}
    {\textit{
Eduardo is an analyst responsible for compiling monthly reports on corporate sales and growth. He uses standard business intelligence workbooks (like PowerBI, Tableau, or Looker) to conduct his analysis.  He also wants to get a sense of how others have analyzed the sales data with these tools. To do so, he must search for data in a large data repository. This is a trial and error process as he tries to formulate the right queries to find data, pipelines, and workbooks that are pertinent to him. Once he has the content he needs on hand, he uses it to augment his current analysis, by including data sources or analysis methods that are new to him, but have been explored by others.
}}
\end{quote}

Next, we contrast this analyst end-user to a machine learning engineer seeking to improve the analysts' content discovery process:
\begin{quote}
    {\textit{
Maya is a machine learning engineer supporting the sales analytics team. She notices that analysts are struggling to find useful content and is seeking to develop a solution to help them. She is exploring recommending content to analysts like Eduardo. Maya realizes she can leverage lineage information and other metadata to identify what content the analyst is currently using (e.g., dataset) and then suggest other content pertinent to them. She is exploring the use of GNNs to address this problem, but needs a tool to understand the utility of its recommendations.}
}
\end{quote}

Our collaborators and the challenges we seek to address through our research are aligned with Maya's persona. However, Eduardo's scenario is pertinent in providing a problem that Maya needs to solve. Moreover, Eduardo is the ultimate beneficiary of this solution and can be consulted about its efficacy, but typically after Maya has some reasonable solution to offer. Our goal is to help Maya understand the quality of recommendations her approach creates, so that she may vet and refine them prior to presenting a solution to Eduardo. Developing a single solution that satisfies both persona's needs is possible but sub-optimal as they have different skills and expectations.

\subsection{Graph Neural Networks (GNNs)}
\begin{figure}[ht!]
    \centering
    \includegraphics[width=\linewidth]{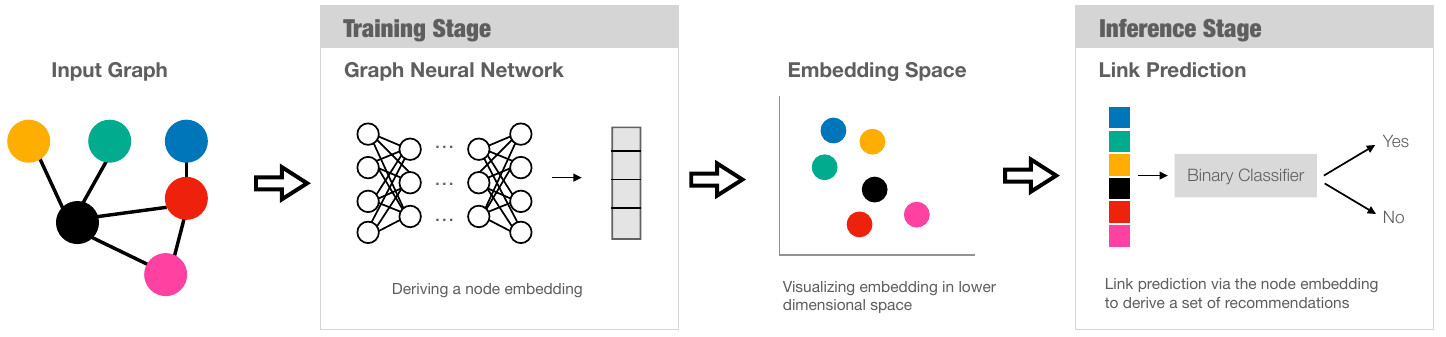}
    \caption{A General Overview of a Graph Neural Network (GNN) pipeline.}
    \label{fig:gnn_pipeline}
\end{figure}

There exist many different machine learning approaches for analyzing graph content, but they come with different trade-offs~\cite{chami:ml_on_graph:2022}. The main limitation of prior machine learning approaches is that they analyze \textit{either} the graph structure \textit{or} features derived from the graph and its nodes, but do not analyze both together -- thus losing potentially valuable information in the model training and inferences steps. GNNs overcome this limitation and can leverage graph structure and features together for different downstream tasks, including recommendation~\cite{hamilton:gnnOG:2017,daigavane:CNN-GNNunderstanding:2021}. 

Similarly, content recommendation approaches have long existed and broadly encompass three types of techniques~\cite{KonstanTerveen:HumanCenteredRecommenderSystems:2021,Jurovac:InteractingWithRecommendersOverview:2017}: content filtering (recommendations based on a person's prior preferences)~\cite{Lops:ContentRec:2011}, collaborative filtering (recommended content based on people's collective preferences)~\cite{Schafer:CFRec:2007}, and hybrid approaches (a combination of both approaches)~\cite{Burke:HybridRec:2002}. As many recommendation tasks can be distilled into graph structures, the use of GNNs has gained popularity over both traditional machine learning and recommendation system approaches~\cite{Wu:GNNRec_Overview:2022}.

\subsubsection{Developing a GNN for Content Recommendation}
We worked together with our collaborators to develop a GNN 
using the two staged pipeline shown in Figure~\ref{fig:gnn_pipeline}. To understand the kinds of data that GNNs ingest and also produce we provide a brief overview of our model training and inference processes.  For more detailed reviews of GNNs we refer to prior work~\cite{Zhou:GNNOverview:2018,Wu:GNNOverview:2021,sanchez:GNNGentle:2021,Zhang:CGNN:2019,daigavane:CNN-GNNunderstanding:2021}. An important item to be aware of is the distinction between the \textbf{input graph}, which is the graph data ingested by the GNN, and the \textbf{dataflow graph}, which describes the architecture of the GNN (and other types of Neural Networks~\cite{Kanit:tensorboard:2018}).  

\vspace{1mm}
\noindent\textbf{Input Data \& Feature Derivation.}~GNNs take graph data as input; other types of neural networks, including those popularly used for image and text data, can only utilize graph input if it is considerably pre-processed, which results in information loss or inefficiency~\cite{daigavane:CNN-GNNunderstanding:2021,Gori:GNN_OG:2005}. The input graph provided to a GNN can take on a variety of forms, including, but not limited to, chemical structures, traffic flow patterns, and lineage data. GNNs can take a single graph as input (e.g., a social network) or multiple graphs (e.g., chemical compounds).

We use the lineage data that is automatically captured by a data catalog system as a singular input graph to a GNN. In this graph nodes represent individual content items belonging to different content types (e.g., database, workflow, workbook) and edges represent known lineages. We use both lineage data and other sources of metadata (usage telemetry, content properties) to derive a set of features corresponding to a node. These features included the type of content, community clustering, centrality, and node degree. 

\vspace{1mm}
\noindent\textbf{GNN Training \& Inference.}
The input graph along with a corresponding set of features is provided to the GNN, where it goes through a standard training procedure (see~\cite{sanchez:GNNGentle:2021} for model training overview). Exploring different GNN architectures (e.g, number of layers, types of layers, etc.), optimization of hyperparameters, and also representation of the features are all part of model training. The output of the training is a node embedding in a high-dimensional space; in simple terms, the initial input graph and features are transformed into an NxM matrix, where N is the total number of nodes and M is the embedding dimensions. We use the node embedding produced by the GNN to perform \textit{link prediction} between any two nodes in the graph~\cite{Zhang:linkprediction:2018}.  Importantly, two nodes can constitute any analytic asset type. For example, a link can be predicted between a user node and an analysis node, or an analysis workbook node and a database. For simplicity, we refer to an initial node of interest as the \textit{source node} and recommendations are derived for one or more \textit{destination nodes}. The link prediction step also produces a probability that any two nodes are connected that can be used to make recommendations.

Understanding the quality of the GNN's recommendations is complex. There is no ground truth data beyond the immediate properties of the graph, but, it is possible to make useful assessments based on these properties. For example, nodes that are directly connected by an edge should have high link prediction probabilities, but it is not clear that those which are more distantly connected, or not connected at all, should have lower probabilities. For example, two distant nodes may belong to the same community, disconnected nodes may share common features. Thus, relying primarily on the topological structure of the input graph as a proxy of quality can limit assessments of the GNN's recommendations. \textbf{To assess the quality of recommendations, MLEs, and SEs need to analyze the recommendations and apply their subjective knowledge and impressions about the input graph.}

\vspace{.5em}
\section{Related Work}\label{related_work}

\subsection{GNN Visualization}\label{rel:gnn_vis}
Existing techniques for visualizing GNNs focus on analyzing the input graph together with quantitative metrics of its performance, typically on prediction tasks. Many of these systems derive an intermediary graph representation to support the assessment of what the GNN has learned. CorGIE~\cite{liu:corgie:2022}, GNNLens~\cite{Jin:GNNLens:2022}, BiaScope~\cite{Rissaki:BiaScopeVU:2022}, and GEMVis~\cite{Chen:GemVis:20222} contrast visualizations of the input graphs global topology against the GNN-derived embedding (or latent space), enabling the end-user to explore these relationships with multiple coordinated views. Embeddings are visualized using dimensionality reduction techniques, such as, UMAP~\cite{McInnes:umap:2018} and t-SNE~\cite{vandermaaten:tsne:2008}. While input graphs can benefit from existing techniques to visualize and interact with multivariate network graphs~\cite{Nobre:graph_survey:2019,vonLandesberger:MVN_survey:2011} we note these approaches have not been widely used in GNN visualization systems.
Node features are also visualized in varying ways and presented concurrently with the input graph and embedding. 
CorGIE~\cite{liu:corgie:2022} and GNNLens~\cite{Jin:GNNLens:2022} also visualize node specific features and an intermediary k-hop topology representation, but vary in their specific treatment of both. While CorGIE, GNNLens, and GEMVis are largely aimed at supporting error analysis, BiaScope focuses more on identifying possible biases in GNN learning.  A  different set of visual design patterns is presented in a design study by Wang~\textit{et al.}~\cite{wang:GNN-XAI:2022}, which uses a visualization of the GNN embedding to drive interactions with specific graph paths. Unlike prior systems, their approach considers the heterogeneity of the GNN's nodes and leverages these properties to create path explanations supported by their novel MetaMatrix view.  These prior approaches are intended for stakeholders with ML/AI expertise intending to refine a GNN model -- an exception being Wang~\textit{et al.}~\cite{wang:GNN-XAI:2022}, which prioritizes scientific domain experts. 

The ML community has developed techniques for evaluating, explaining, and visualizing GNN behavior~\cite{Pope:GNN-ExplainabilityMF:2019,Shchur:GNN-EvalPitfalls:2018,Yuan:GNN-taxonomy:2022}. 
For example, GNNExplainer~\cite{Ying:GNN-Explainer:2019}, CF-GNNExplainer~\cite{Lucic:CF-GNNExplainer:2022}, XGNN~\cite{Yuan:XGNN:2020}, and SubgraphX~\cite{Yuan:Subnetwork_XAI:2021} methods both use visualizations of the overall input subgraph to show the influence of specific nodes and neighborhoods on a GNN's prediction. They also visualize the features of individual nodes. The GNNLime~\cite{Huang:GraphLIme:2022} and GraphSVX~\cite{Duval:GNNSVX:2021} techniques optimize the LIME and SHAP approaches to generate and visualize explanations. As GNN research continues we anticipate that additional encoding and interaction techniques will be developed, either independently of, or with collaboration with, the VIS research community.

We draw on common and emerging design patterns from prior work in developing~\sys. 
We also examine the applicability of these approaches to recommendation tasks and explore alternative designs. As with the majority of prior research we target ML/AI domain experts, specifically MLEs and SEs. 

\subsection{Visualizations for Neural Networks}\label{rel:NN}
Techniques for visualizing Neural Networks (NN) more generally are pertinent to our research as well. Hohman~\textit{et al.}~\cite{homan:VisDLsurvey:2019} summarizes the role of visual analytics in Deep Neural Networks (DNNs) as a set of interrogative questions that are then tied to specific visual encodings and their uses. While DNNs vary in their applications and architectures~\cite{Liu:DNN_Arch:2017}, there are overlaps with GNNs, particularly towards visualization of embedding and latent spaces (e.g.,~\cite{Boggust:EmbeddingComp:2022,Liu:LatentSpaceCart:2019}), dataflow graphs (e.g.,~\cite{Kahng:ActiVis:2018, Kanit:tensorboard:2018, Wang:CNN_explore:2021}), and features (e.g.,~\cite{olah:NN_features:2018,homan:summit:2019}). It is also noteworthy that many of the NN visualization techniques and toolings are rooted in more general research concerning visual analytics for machine learning~\cite{Yuan2021ASO,Scha:Vis4ML:2019}. We do not extensively survey individual visual analytic approaches for NN here, as they are less pertinent to our examination of GNN-based recommendations. However, we acknowledge the influence of these prior works on GNN visualization.

\subsection{Visualization for Interactive Recommender Systems}\label{rec:vis}
User interactions with recommender systems have been explored by the visualization and human-computer interaction (HCI) communities~\cite{Valdez:RecHCI:2016,Huaiqing:VisRec:2016}. User interfaces of these prior efforts have been motivated by purposes: 1) eliciting information from end-users via interactions and 2) present the results of recommendations for end-users to explore~\cite{Jurovac:InteractingWithRecommendersOverview:2017}. Our research is aligned with the second motivation.

Visualizations for recommender systems specifically, play an important role in explaining recommendations, allowing end-users to fine-tune and control their preferences, and take next actions.  PeerChooser~\cite{ODonovan:PeerChooser:2008} visualizes recommendations as a node-link diagram and invites user interaction to interrogate the results. These systems are early examples of visualization for recommendation, the prior majority (e.g., MovieLens~\cite{Miller:MovieLens:2003}, GroupLens~\cite{Resnikc:GroupLens:1994}) prioritized ranked lists~\cite{Jurovac:InteractingWithRecommendersOverview:2017}. Systems like TasteWeights~\cite{Bostandjiev:TasteWeights:2012}, TalkExplorer~\cite{verbert:TalkExplorerRec:2013}, and SetFusion~\cite{Parra:SetFusion:2014} use a combination of node-link diagrams, clusters (TalkExplorer), and Venn diagrams (SetFusion) to visualize the diversity of recommendations and overlap of preference. These systems also provide widgets allowing end-users to weight different factors in recommendation and visualize their effects. In an alternative approach, VizCommender~\cite{Opperman:VizCommender:2021} explores different techniques for aligning recommendations with human judgements in content-based recommendations of data assets and presents the results as a visual gallery. 
There are also techniques for visualizing individual user recommendations through radar  or scatter plots of preferences~(e.g., ~\cite{Millecamp:ControllingSpotifyRecommendations:2018, Tsai:ExploringSocialRecommendations:2019}). 

We observe some overlap between visualization approaches from recommender system and GNNs, but also note that much of this work is siloed into specific application domains and disciplinary subsets. Our research takes an interdisciplinary lens, balancing design trade-offs  between the design spaces of GNNs and recommender systems.
\vspace{.5em}
\section{Data and Task Abstraction}
Our collaborators' data and tasks are unique to their domain goals. We briefly provide an overview of the input data graph and GNN that are used to generate analytic asset recommendations and are visualized by~\sys. To make our findings transferable to other domains and applications, we present their data and tasks in abstracted terms.

\subsection{Data Abstraction}\label{data_overview}

\begin{figure}[h]
    \centering
    \includegraphics[width=\linewidth]{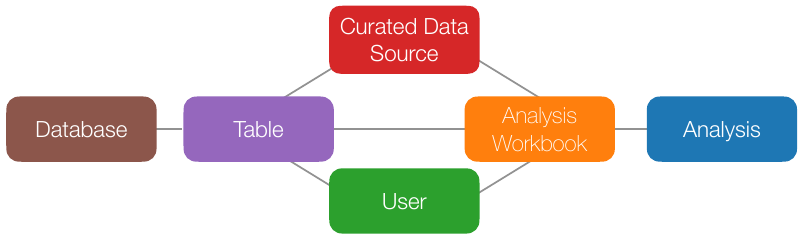}
    \caption{Simplified representation of the input graph of analytic content for recommendation. We show the types of analytic assets that are recommended, including user information, as well as the lineage relationships between assets and (when applicable) users. }
    \label{fig:input_grapj}
\end{figure}
Our collaborators capture end-user analysis through a data catalog and corresponding data repository (see Section~\ref{intro} and~\ref{analytic_content} for a general overview). The catalog graph stores metadata for different content types 
and also encodes the asset lineage that captures the relationships between assets. 
For example, a dataset is a type of content that can form a lineage relationship with one or more analysis workbooks. Other data sources included telemetry and usage data that identity when and by whom analytic assets were created and viewed respectively. %
The combination of these two data sources can be used to produce personalized recommendations for analyst end-users. %
This \textbf{input graph} was derived from our collaborators' test site and consists of 
211,515 nodes and 459,150 edges. We have obfuscated the input graph for presentation in this manuscript.

We can abstract our collaborators' data into the following types:

\begin{itemize}
    \setlength\itemsep{-0.1em}
    \item \textbf{A multivariate network} containing derived feature attributes for each node. Each node in our network has an asset type~(\autoref{fig:input_grapj}) and edges are connections between one or more asset types.
    \item \textbf{A dataflow graph} that describes the architecture of the GNN and the computations used to train it. 
    \item \textbf{An embedding} that maps the nodes and structure of the input graph into a lower-dimensional vector representation. 
    \item \textbf{A tabular dataset} comprising the features of nodes and their probability of being linked to a source node.
\end{itemize}

In developing \sys, we consider these different abstract data types collectively with the design space cross-overs between GNNs and recommender systems.

\subsection{Task Abstraction}\label{task_abs}
Through iterations and discussion with our collaborators, we determined that our collaborators visual analysis goals were to contextualize recommendation results and qualitatively assess their validity. We can further breakdown these visual analysis goals into the following tasks:

\begin{itemize}
    \setlength\itemsep{-0.1em}
    \item \textbf{T1 Summarize} the GNN's recommendation results for a given input node, including its relationship to other nodes (e.g., correlations and outliers), features, and prediction probabilities.
    \item \textbf{T2 Compare} recommendations for a single node. For example, comparing recommendations according to their distance from the node of interest.
    \item \textbf{T3 Contextualize} the results for a single input node by understanding the types of assets that were recommended, the distribution of their probabilities, and the similarities of the features.
    \item \textbf{T4 Validate} the GNN's results  by applying domain knowledge to accept or reject the recommendations.
\end{itemize}

Through discussion with our collaborators, we determined it was more concrete to examine a single node at a time together with the subset of its recommendations. Collaborators often had specific questions pertinent to specific nodes and visualizing the full spectrum of nodes and links (existing and predicted) was less relevant.

\vspace{.5em}
\section{\sys}\label{reckomgnn_interface}
We developed \sys through an iterative design process~\cite{sedlmair:dsm:2012}, beginning with paper and pen-based iterations, before reifying these ideas into a Web-based application that we continued to refine with our collaborators' input.

The final interface consists of three panels~(\autoref{fig:teaser}): the data panel, the recommendation overview panel, and the recommendation detail panel. The primary users of \sys are MLEs and SEs that are seeking to interrogate the GNN results.

\noindent\textbf{Data Panel}. Given some source node in the input graph, the data panel allows the MLEs and SEs (our target users) to \texttt{summarize(T1)} all recommendations to other destination nodes, including the prediction probability, and to record a qualitative assessment of  the prediction quality. Features and other properties of destination nodes are shown to help users make their assessments.  Users also often have questions about specific source-destination node combinations, for example, if they feel strongly that some data source (destination node) should be recommended to an analyst because they are using a related dataset in their present analysis (source node). Further filtering the table to a destination node of interest can \texttt{validate(T4)} their hypothesis. 

\begin{figure}[h!]
    \centering
    \includegraphics[width=\linewidth]{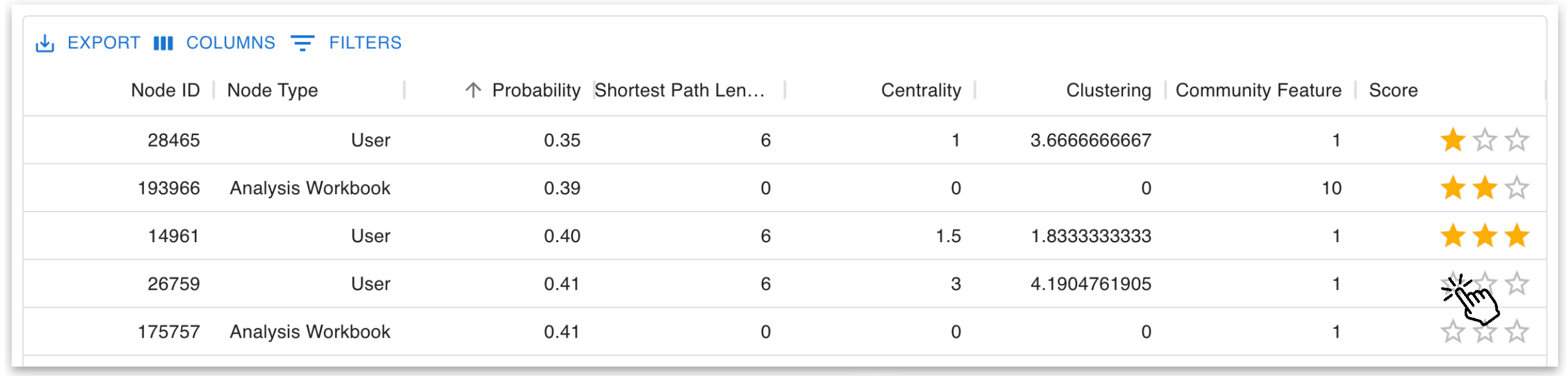}
    \caption{Users can score the quality of a recommendation in the table view of the data panel. These annotations are retained and can be exported for further analysis in an existing DS/ML workflow.}
    \label{fig:annotation}
\end{figure}

The data panel is supported by a node selection widget and a table view. The node selection widget enables the user to select a source node from the input graph (via its node id) and displays some of its basic properties. The table view displays the destination nodes of the predicted links. As  the user examines the GNN's recommendations through the three components, they can log their qualitative assessments in the table view (\autoref{fig:annotation}).  By allowing users to annotate recommendations, they can bring their subjective expertise to bear on the quality of the results and not rely solely on quantitative metrics. These annotations are stored and can be modified at any point, they can also be exported as a CSV to a programmatic environment (e.g., Python, R), where they can be further explored. As MLEs and SEs update the GNN, they can import the results to~\sys for further analysis and refinement. Importantly, MLEs and SEs expressed a preference for this type of import-export integration in order to allow for easier incorporation into existing, and ever changing, workflows.

\noindent\textbf{Recommendation Overview Panel}. While the data panel provides a summary, the recommendation overview panel makes it easier to  \texttt{compare(T2)} recommended nodes simultaneously. The recommendations are shown to users along with their predictive probabilities and node properties. Users can visually examine and interact in order to develop insights into how the GNN may be making recommendations, and make assessments of the result quality. For example, users may wish to consider the relationship between the distributions of prediction probabilities and the path length between the source nodes and possible destination nodes. They might be interested in finding nodes with high probabilities that are distant in the input graph and wish to examine this further. Notably, this kind of exploration does not require visualizing the input graph directly, as prior work~(Section~\ref{rel:gnn_vis}) has emphasized, but rather its derivative properties.

The recommendation overview comprises a probability distribution panel, broken down by node type, and a recommendation attribute panel that provides further details of the relationship between the probabilities and attributes from the input graph. The recommendation probability coordinates the views in the two panels, helping the end user \texttt{compare(T2)} recommendations.

\noindent\textbf{Recommendation Detail Panel}. Users can drill deeper into a specific set of recommendations to further \texttt{contextualize(T3)} individual factors that contributed to the GNN's recommendation and \texttt{compare(T2)} to similar nodes in the embedding space.  The views of the detail panel are complementary to the overview and data panels, and offer an alternative and deeper level of granularity to examine recommendations.

\vspace{.5em}
\section{RekomGNN Design}\label{reckomgnn_design}
We now describe the design choices of the visual encodings and interaction techniques that support the panels of \sys's interface. We discuss the rationale of these choices in the context of our data and task abstractions and with respect to trade-offs between the GNN and Recommender system design spaces.


\subsection{Visual Encodings}
We break down the visual encoding design choices for the panels of the interface components. A design constraint for all of these views was the size of the input graph~(Section~\ref{data_overview}), which made it challenging to display all of recommendations for a source node and requires aggregating or sampling the data. We chose a stratified sampling approach that draws a representative sample from across the range of recommendation probabilities. Repeated draws would yield new recommendations to interrogate.

\subsubsection{Data Panel}
The data panel allows users to inspect the underlying \textbf{tabular dataset} of the GNN's recommendations. Through simple interactions they can choose the node features to inspect -- ordering, or filtering the rows by the values of different columns, and even removing columns they might not be interested in exploring.  As previously indicated (Section~\ref{reckomgnn_interface}), this view also allows users to score the recommendation quality. Finally, the \textbf{table} view of the data panel is also consistent with how MLEs and SEs regularly interact with their data and as such was an important anchor in their analysis ahead of proceeding further with a visual analysis. 

The \textbf{table} view contains information not only about the source and destination nodes, but also derived properties of the \textbf{input graph} such as the shortest path between two nodes and communities within the graph. We refer to \textit{node features} as existing (e.g., content type) or derived (e.g., centrality) attributes for a single node, whereas as \textit{graph attributes} are derived from the relationship between multiple nodes (e.g., shortest path). The design choice of having node features and graph attributes in a table is a departure from prior systems that visualize GNNs (Section~\ref{rel:gnn_vis}) and that visualize the input graph or a derived intermediary graph (e.g, k-hop topology). We elaborate on this choice in Section~\ref{design-rationale}.

\subsubsection{Recommendation Overview Panel}\label{rec_overview_desigh}
The overview component visualizes the recommendation distributions of probabilities together with node features and input graph attributes.  With this set of complementary views the analyst can explore the univariate and bivariate relationships between the recommendation probabilities and the different topological attributes, building up an intuition toward how the input node is related to these recommendations. We argue that it is challenging to explore these kinds of insights in prior systems that visualized GNNs. For example, while GNNLens~\cite{Jin:GNNLens:2022} also visualizes node features they do so primarily to identify influential nodes for their prediction tasks and less so to compare the predictions a single GNN makes; their implementation instead focuses on comparing predictions between different GNN implementations. CorGIE~\cite{liu:corgie:2022} prioritizes the relationships between the input graph topology, embedding, and feature space. While CorGIE does enable some comparison of features, these are based on derived distances rather than the node's features. We enable these comparisons through the \textbf{probability histogram} and \textbf{multi-axis scatter plot} views.

In the \textbf{probability histograms} view analysts can observe the recommendation probability distributions according to the six asset types (\autoref{fig:input_grapj}) of our collaborators' data. These asset types are not fixed, and these categories can be easily modified to different types and quantities of analytic assets and domains (e.g., e-commerce applications). The probability distribution allows users to evaluate the model's confidence with respect to recommending an asset type.  Moreover, this comparison of distribution across asset types enables users to triage potential issues with asset types. For example, if the GNN recommends all database assets with the same probability it may be an indicator that the features of the nodes for database assets are not informative. Unlike prior work, which also uses histograms to visualize feature distributions, we prioritize the recommendation probabilities because of their importance in aiding the understanding of the GNNs results.

\begin{figure}[t]
    \centering
    \includegraphics[width=\linewidth]{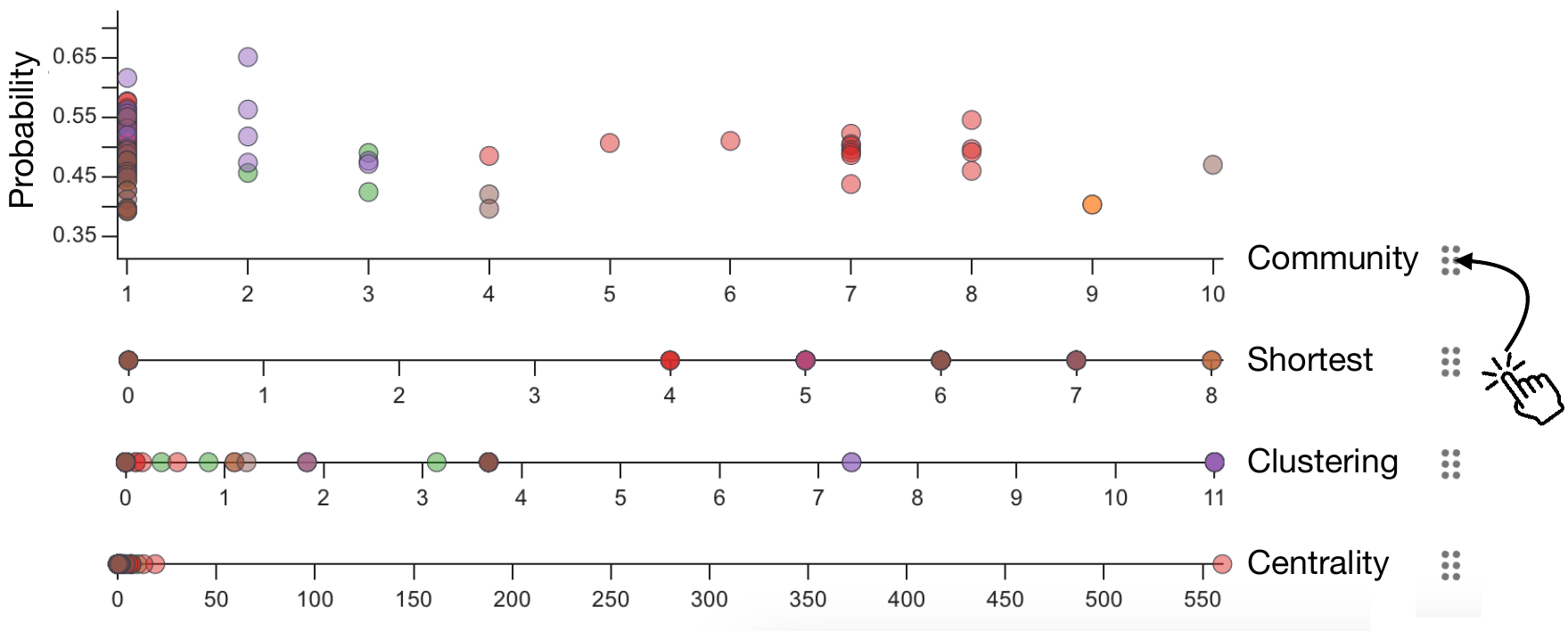}
    \caption{Users can drag-and-drop the x-axis to interrogate the relationship between the recommendation probability and different graph attributes and node features.}
    \label{fig:multi-axis}
\end{figure}

The \textbf{multi-axis scatter plot} (~\autoref{fig:multi-axis}) shows the bivariate relationships between the recommendation probabilities and different node features and graph attributes. While the y-axis remains fixed, the x-axis can be changed through a drag-and-drop interaction. As the x-axis changes, it is possible to visually assess the distribution of probabilities across different asset types as well as features and attributes. In the example shown in~\autoref{fig:multi-axis}, it is possible to see that regardless of the community they occur in, user nodes (green color) are consistently predicted with lower probabilities compared to other asset types. By shifting the axis to different features, users can establish factors that might be influencing these results. Prior systems have used feature matrices~\cite{wang:GNN-XAI:2022,Jin:GNNLens:2022,liu:corgie:2022} or feature strips~\cite{liu:corgie:2022} to show similar insights. We chose the alternative of the interactive multiaxis scatter plot, over the more static feature matrices, because it was easier to see and interrogate bivariate relationships; dynamically changing the x-axis amounted to probing the GNN's results in order to surface insights that could be further contextualized or validated. 

\subsubsection{Recommendation Detail Panel}
The detail component is comprised of two adjacent views: a \textbf{probability-attribute rank} and \textbf{embedding projection}.  These views help the user to inspect recommendations they selected in the table or overview components and analyze them in more detail. They provide a complementary perspective to the views in the overview panel by recasting some of the data as different visual encoding, and reducing visual noise by only including the top and bottom five by node category. This design choice is intended to support the user to further \texttt{compare(T2)} and \texttt{contextualize(T3)} recommendations.

\begin{figure}[t]
    \centering
    \includegraphics[width=\linewidth]{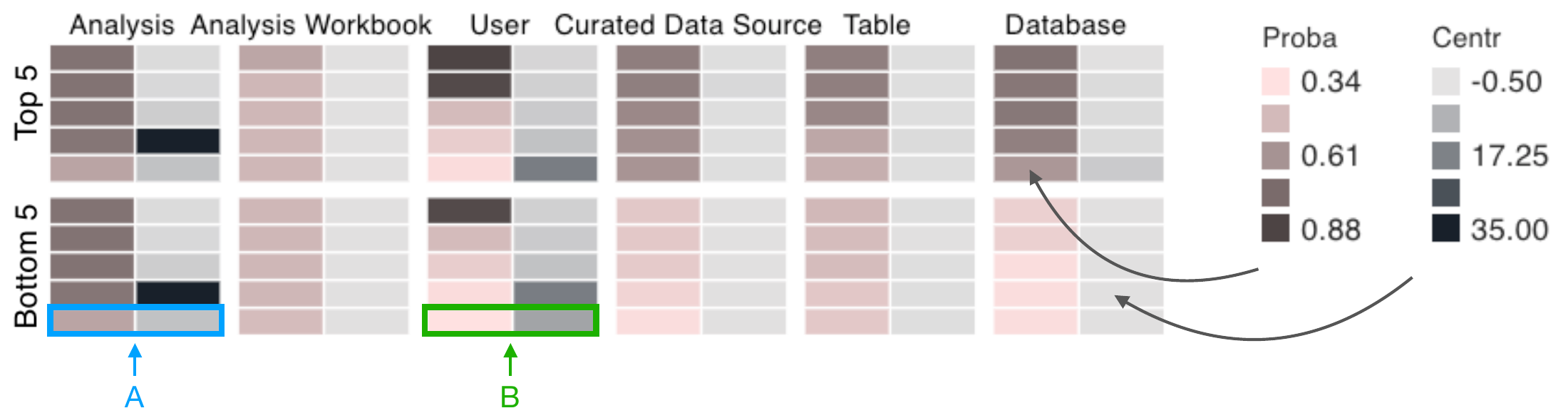}
    \caption{The Probability-Attribute Rank visualizes a recommendation as a pair of rectangles colored by probability and a selected feature respectively. The top and bottom five recommendations by node feature are displayed. We highlight two recommendations (A and B) with similar centrality but different recommendation probabilities.}
    \label{fig:array}
\end{figure}

The \textbf{probability-attribute rank} view~(\autoref{fig:array}) is comparable to the feature matrices and stripes in prior systems~\cite{liu:corgie:2022,Ying:GNN-Explainer:2019} as well as the multiaxis scatter plot. However, instead of showing the probability against multiple features and attributes, the probability-attribute rank displays the probability and selected feature as a rectangle pair. Each rectangle pair corresponds to one recommendation and comprises a recommendation probability (left rectangle) and a feature (right rectangle). The recommendations are sorted by probability, and only the top and bottom five are displayed. This allows one to quickly differentiate between top and bottom-ranking predictions. Two divergent color palettes are used to \texttt{compare(T2)} the probabilities and features, with darker colors of both scales representing higher values (e.g., higher probability, higher centrality, or longer path length). This design choice makes it easier to identify interesting bivariate relationships.

\begin{figure}[b]
    \centering
    \includegraphics[width=\linewidth]{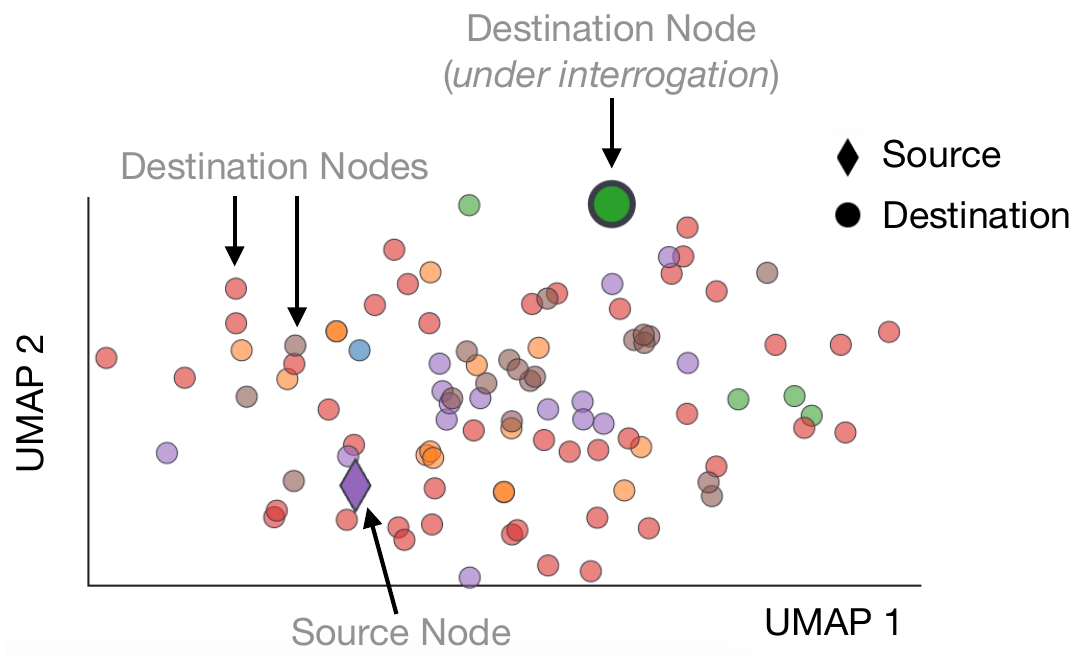}
    \caption{Embedding projection showing source and destination nodes. Larger destination nodes are those under interrogation in other views and are interactively linked via hovering interactions.}
    \label{fig:embedding}
\end{figure}

We retain the \textbf{embedding projection} view~(\autoref{fig:embedding})  that is a standard in prior GNN visualization tools as well as other neural networks in general (Section~\ref{related_work}). This view projects the node embeddings produced by the GNN~(\autoref{fig:gnn_pipeline}) into a two dimensional space using UMAP~\cite{McInnes:umap:2018}. We use shape to differentiate between source ($\diamond$) and destination ($\circ$) nodes. The points are also colored according to asset type. While other views highlight univariate distributions or bivariate relationships of different features with the probability, the embedding view represents a multivariate summary of the data. Users can further \texttt{contextualize(T3)} recommendations by using the distance between points as a proxy for the similarity for the nodes. This too can be used to triage the recommendations of the GNN. For example, if the source node is a database and the proximal other database destination nodes this could also be a signal of an uninformative feature space for the nodes. The embedding projection could also be used to identify outliers, which was more difficult to assess in other views.

\subsection{View Layout and Coordination}
\begin{figure}[h!]
    \centering
    \includegraphics[width=\linewidth]{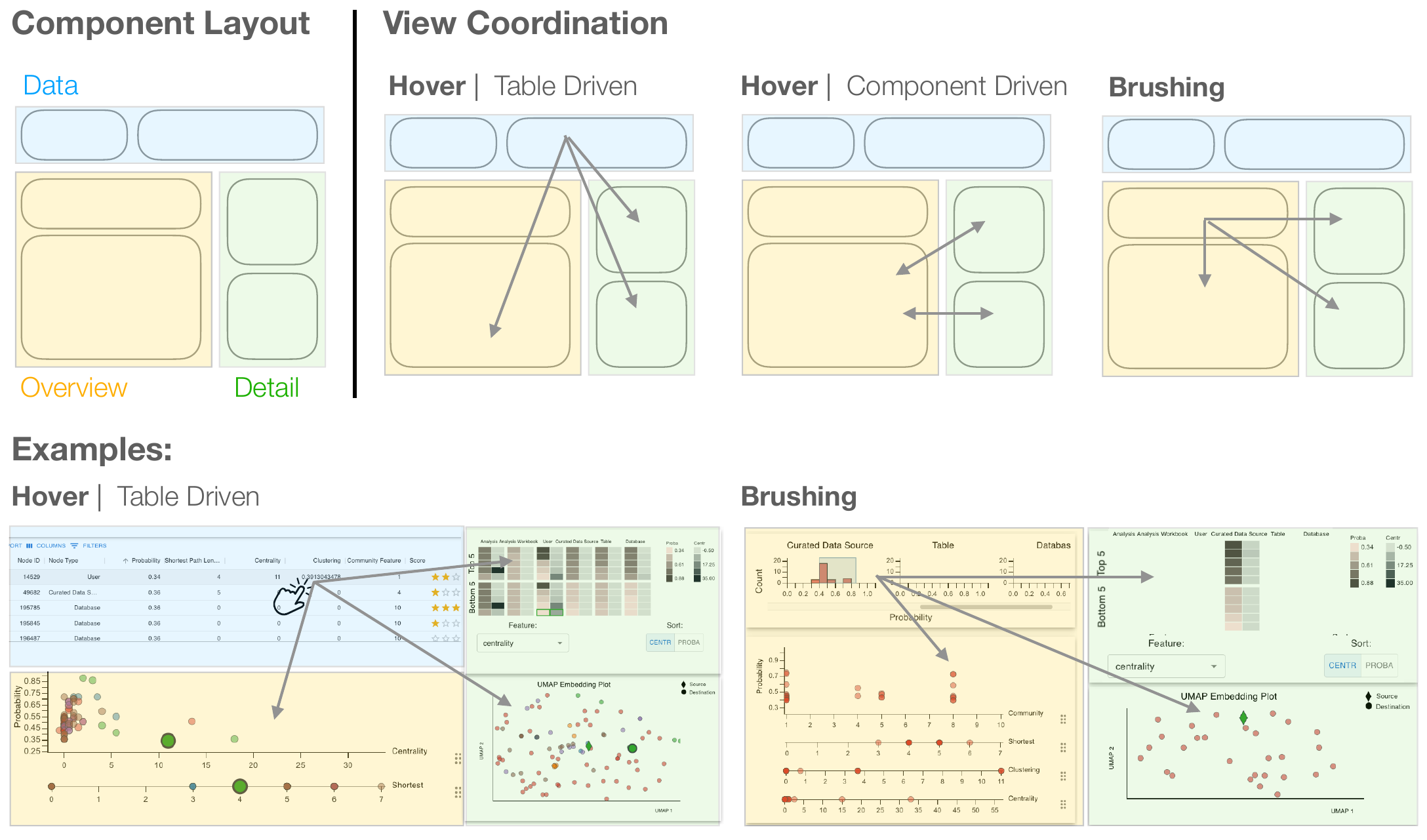}
    \caption{View coordination is achieved primary through hovering and brushing interactions and has effects across panels.}
    \label{fig:view_coordination}
\end{figure}

Views are grouped within the table, overview, and detail components and are linked through \textbf{hovering} and \textbf{brushing} interactions (\autoref{fig:view_coordination}), which serve to emphasize and filter nodes, respectively, across views.

\textbf{Hover} interactions can be initiated either from the table or through different components to visually emphasize individual destination nodes from a source node. Hover interactions that are initiated from the table have effects on views within other panels. However, interactions initiated from components \textit{do not} act to filter the table. The consistency in the table view anchors the user's visual analysis experience, modifying it in response to panel interactions was disorienting.

In the scatter plot and embedding views, nodes are visually emphasized by changing their size relative to the others in the plot. We also experimented with changing both the size and opacity of nodes to highlight the destination node of interest. However, the size of the dataset resulted in a varied opacity across the plots because some regions were more dense (i.e., contained more data points) than others, which was more distracting than it was helpful. Keeping the opacity consistent and modifying the point size only was less disorienting as fewer aspects of the plot were changed; the consistency in the views in response to hover interactions made it easier to identify the emphasized node.

\textbf{Brushing} cross-filters the data displayed in the linked views. This interaction can be initiated within the probability histogram and multi-axis scatterplot views and enables users to filter to specific node types within a probability range of interest. For example, a user may want to interrogate database assets that receive low probabilities in the GNN's recommendations to assess what factors contributed to these results. 

\subsection{Design Rationale}\label{design-rationale}
Our design choices diverge from existing design patterns for GNN visualization and also draw upon the design space from recommender systems. While we have already alluded to some of these trade-offs, we now discuss these differences relative to prior work.  

\noindent{\textbf{Why not visualize the input graph?}} Prior visual analytics tools and recommender systems feature the input graph as a central component. 

In contrast, we chose not to display the input graph and instead visualized derived properties of the graph, such as communities and path lengths. Our primary rationale was that through our study we did not find evidence that the input graph was helpful for making validity assessments of the GNN's recommendations.

For example, if a user observes a high recommendation probability they could examine the input graph to see how distant two nodes are. This action is equivalent to mentally calculating the shortest path between two nodes and in our approach, we make this calculation for the users and display its result.
Our secondary rationale for our choice is that the size of the input graph made it undesirable to examine even if we had ultimately chosen to include it. Prior work does not discuss how well their techniques scale to large graphs with heterogeneous nodes. While this poses an interesting research problem, it has limited utility for our collaboration's domain goals.

\noindent{\textbf{Why not visualize the data flow graph?}} Prior systems for the visual analysis of neural networks have shown the models architecture as a data flow graph (Section~\ref{rel:NN}). While we have leveraged other aspects of NN visualization, for example embedding projection and alternative ways to visualize features, there was limited benefit to also showing the GNN's architecture as a data flow graph. Prior visualization systems interrogating the results of GNNs for different tasks also do not visualize the data flow graph.

While the data flow graph can be useful to refine a GNN after interrogating its results, tools already exist in pytorch~\cite{pytorch:2019} and tensorflow, two widely used NN libraries, to do so. MLEs prefer to perform these refinement tasks within their existing workflows and as such including the data flow graph in~\sys had limited utility. 

\noindent{\textbf{Why emphasize the recommendation probability?}} The link prediction probability, or recommendation probability, is a central feature of \sys, that enables us to sample the data as well as link and coordinates its views. There are other node features or graph attributes that we could have used in lieu of, or allowed to be alternated with, the recommendation probability. However, for the more general recommendation tasks that drives our collaborators domain goals, the probability is the primary mechanism by which analysts would eventually be served recommendation. Tools for interrogating recommender systems (Section~\ref{rec:vis}) also prioritize visualizing data that an end-user is likely to actually be served, similar to our choice of emphasizing the recommendation probability. Moreover, these recommender systems aim to actively solicit feedback on the quality of these recommendations, as is possible in \sys's data panel. In contrast, GNN visual analysis tools are not as opinionated about the features or attributes that are prioritized  and do not support user annotations of data quality.

\noindent{\textbf{Why subsample the data?}} The size of our collaborators input graph necessitated trading-off between approaches to visually aggregate the data or sample it. Aggregating  introduces a coarseness to the visual display that can obscure trends and outliers that are of interest to users. For graph data in particular it can be challenging to condense a complex graph topology, and even though there exist techniques to visualize dense graphs, MLEs and SEs can struggle to interpret them. Sampling allows users to observe individual data points, but also omits much of the data. Compared to aggregation we found that sampling, and if necessary repeated sampling, provided the appropriate level of granularity to examine the GNN's recommendation results. Importantly, our design choices to forego displaying the input graph and to prioritize the recommendation probability made sampling the data feasible and accurate, while still proving useful to our collaborators goals. 

\vspace{.5em}
\section{Implementation}
\sys is implemented in JavaScript using React.js, D3, and visx. The GNN model was implemented in Python using Pytorch~\cite{pytorch:2019} and the PyG~\cite{pyg:2019} libraries. Graph data are stored in a propitiatory data catalog from which we derived a simpler subgraph that we store and analyze entirely using NetworkX~\cite{hagberg:network:2008}. 
\vspace{.5em}
\section{Usage Scenarios}
We validate the utility of \sys in two forms. First, we present usage scenarios detailing how users can realize their tasks using our tool. Second, we provide usability feedback from our MLE and SE collaborators, which we present in the subsequent section. When discussing the recommendations made by the GNN, we use the terminology that reflects the structure of the input graph. For example, recommendations are always initiated by first selecting a node from within the graph from which to explore recommendations. Most often, this source node represents a user, but, it can also represent other content types, such as a dataset, workbook, etc. In the former, recommendations are based on the user alone.  While in the latter context, recommendations are made based on content the user has on hand, for example, a notebook the user is presently working with; the notebook serves as the source node. 

\subsection{Assessing Recommendations with Expert Knowledge}

Ensuring personalized recommendations that cater to the specific needs of users is crucial in business intelligence organizations. Analysts, data scientists, and software engineers primarily seek recommendations for data-related assets such as datasets, data sources, and tables to support their analysis.  This usage scenario considers how a machine learning engineer can use ReckomGNN in combination with their expert knowledge to assess the quality of different content recommendations by examining the probability distributions of different content types.

Maya, an MLE,  begins her analysis by selecting a source node that corresponds to a particular user to see the types of content that are recommended to them. In the data catalog graph, these users are analysts who either own content (e.g., created a workbook or dataset) or have viewed some content. From this initial selection, she can explore the probability distribution histogram to assess not only the types of recommendations but, also the GNNs' confidence. She uses a cut-off of $p > 0.5$ as an indicator of whether a certain type of content is recommended. From this overview, she can \texttt{compare (T2)} how likely it is that the GNN would recommend some content.

First, Maya examines the workbook recommendations that the GNN makes. She notes that the range of probabilities is between 0 and 0.4 and that none would reach the recommendation threshold of 0.5. Using the table view, she looks up an analytic workbook she knows was created by the user and should have a higher recommendation probability, she notes the low probability score and flags it by selecting one star to indicate a poor quality result~\texttt{(T4)}. To triage the problem, she considers the different features in the multi-axis scatter plot, she emphasizes the workbooks by brushing over the histograms of probability distributions. From this action and sees that there is not much differentiation among the features, which may impact how the GNN is making recommendations. She makes a note to return to the feature generation and pre-processing as one way to improve recommendations.

Overall, Maya sees that for this particular user, the recommendations are quite sparse -- the GNN doesn't appear to have enough information to make good recommendations. She can contrast this to a different user node~\texttt{(T2)} where the probability distributions have a wider range for different content types. Through the comparison of multiple nodes, she attempts to uncover the different factors influencing the recommendations. She takes these insights back into the development process to continuously refine and tune the GNN. She uses the recommendation she's flagged in the table view as a way to return to and monitor if and how the GNN's recommendations change over time.

\subsection{Exploring Recommendation Patterns and Diversity}
In the second usage scenario, Maya explores patterns of recommendations by comparing different features and attributes associated with recommended nodes. Her goal is to understand how these patterns influence the diversity of recommendations generated by the GNN. She is particularly interested in understanding whether there is an imbalance in recommendations (e.g., one type of content is recommended more frequently than others) and if so, identify ways to correct this issue.

\subsubsection{Discovering Trends within the Overview Panel}

Maya begins her analysis by using the node selection widget of RekomGNN. Once again,  she selects a recommendation source node she knows pertains to a user to begin her investigations. She looks at the probability histogram pertaining to the users, those with higher probabilities suggest other users that are similar to the one she is currently investigating - she wants to \texttt{compare (T2)} them to get a sense of what makes those users similar or different to the source node she selected. She can do so by brushing over the histogram, which highlights different attributes across the interface panels. In the multi-axis scatter plot, this action reveals correlations between the wide range of probabilities and the node and graph attributes. She swaps the axes of the scatter plot through the drag-and-drop functionality to uncover these correlations, stopping when she notes the positive correlation between
probability and node centrality. Nodes with higher centrality, indicating greater importance or prominence in the network, receive higher recommendations from the GNN. This correlation validates the GNN's performance when compared to a simple popularity baseline~\texttt{(T4)}. Maya verifies this correlation pattern for multiple user nodes, revealing that  the GNN recommends popular content with higher probability.

\subsubsection{Probability diversity within node types}

Maya conducts a deep dive into curated data sources, which exhibit a bell-shaped probability distribution centered at 0.5 for most user source nodes. By exploring multiple source nodes, brushing over different probability ranges in the probability histogram, and examining the multi-axis scatterplot, Maya discovers that low-probability recommendations mainly come from specific communities, while high-probability recommendations originate from other communities.
To gain additional insights, she continues to analyze the distributions and communities of the curated data sources to understand why certain communities receive higher or lower probabilities. This analysis may involve exploring the content or attributes of nodes within these communities to uncover any patterns or characteristics that can be leveraged to enhance the GNN's performance. Finally, Maya utilized the probability-attribute rank view to contextualize (\texttt{T3}) the top and bottom five recommendations to be broken down by content types. She sees that the top 5 databases were typically located near the source user node, while the bottom five databases lacked any direct path connecting them to the user. Consequently, databases in separate connected components received considerably lower recommendation probabilities than those within the same component. This outcome deviates from the desired objective of promoting diversity and encouraging users to explore a wider range of databases. Maya assigned a 1-star rating to the bottom five database recommendations within RekomGNN to remember to address this issue. She intends to tackle the diversity challenge outside the tool by considering strategies such as introducing synthetic or generated samples that bridge the gaps between disconnected components.

\vspace{.5em}
\section{Collaborator Feedback}

Throughout the development of~\sys we checked in with our collaborators on a biweekly basis to gather feedback on the evolution of our interface and its functionality. Our collaborators constituted a team of 4 software engineers, 4 machine learning engineers, and one project manager. All individuals had experience developing aspects of search and recommendation systems for analytic content. The use of GNNs for recommendation tasks was new to everyone involved. 

We consulted our collaborators through the iterative design of \sys, and as such, many of their perspectives are already captured in the system's design. For example, the ability to annotate recommendation quality and filter to specific destination nodes arose during design discussions. One collaborator pointed to the value of gathering qualitative insights for verifying the GNN's output: \textit{``one of the things I was thinking about is for qualitative data whether it's possible to also put a destination node and that would make it so that you could actually verify by yourself''}. The value of those annotations to one collaborator was to follow on whether \textit{``this is maybe interesting to be investigated or does is definitely something that as a human [I think] is valid''.} Originally, we had intended to use textual annotations as a way to indicate the quality for the recommendation, but two collaborators felt a numeric indicator would be more useful: \textit{``some kind of scoring even [would be useful and] would be a new way to feedback [on] those annotations''}. As we iterated towards a more finalized interface the design, with was concluded that both numeric scores and some qualitative annotations were still useful because, according to one collaborator, \textit{``actually collecting some kind of qualitative data to help you you know to test it in the future.''}. These collective impressions of the recommendation quality, derived from~\sys, could also be used to triage possible issues and revisit model updates. As such, being able to easily export these qualitative annotations into a programmatic environment was useful. 

Participants also had an overall positive impression of the functionality of the final~\sys interface. All agreed that it was interesting to contextualize the recommendations of the GNN with respect to the distance between nodes. An SE collaborator expressed that \textit{``if you can find insights or like, connections between things that are very distant in the graph, those are things that would never pop up any other way''} and that the ability to probe into these insights with~\sys would \textit{``possibly be super valuable''}. As we indicated earlier, seeing the physical hops between nodes was not necessarily valuable, especially because GNNs can exploit feature information in addition to the input graph's topology. Note also that the input graph can have disconnected components (e.g., analyses that use different datasets) but that nodes in separate components can still be relatable through their features. 

We observed that there exists a learning curve to identify and integrate information across the different panels to verify the recommendation quality. As each individual makes a subjective assessment of quality, it was difficult to identify concrete avenues to lower the burden of this experience. However, taking the feedback together,~\sys was able to provide a holistic view of the GNN's recommendations to perform such assessments.

\vspace{.5em}
\section{Discussion}

Content recommendation is ubiquitous across many domains but has not been previously explored in the context of data analysis. Identifying and using pertinent data that is stored within large heterogeneous repositories remains complex and can benefit from approaches that recommend analytic assets. We collaborated with a team of MLEs and SEs to explore the potential of leveraging the lineage and metadata of these repositories, which form a  large graph, together with newly developed GNN-based approaches for content recommendation. While the ultimate goal of this collaboration was to develop an approach to serve relevant content to analyst end-users, a more immediate and pressing goal was to validate the quality of the GNN's results and establish the viability of this approach. 

We conducted a design study that examined the joint design space of GNN and recommender system visualization to arrive at a set of choices for assessing the results of a GNN recommendation pipeline. We ground these choices in data and task abstractions that we derived through iterative discussions with our collaborators. We reify these choices in the design and development of~\sys.  Importantly, we highlight and discuss our design choices that go against established patterns of prior research. Through the presentation of two usage scenarios and feedback from our collaborators, we demonstrate the utility of~\sys. Our findings have implications for the visualization of GNNs in the context of recommendation tasks. However, the ubiquity of graph data across domains and the growing popularity of GNNs enables our findings to apply broadly. Our research also contributes to the growing body of visual analysis literature for GNNs. 

\subsection{Limitations}
Design studies, by their nature, focus on very specific applications and set of users. As such we heavily lean on our collaborators' semantics and expertise when making choices about visual encodings, their complexity, and the density of the data. While this approach illuminates unique goals, data, and tasks that enrich our overall understanding of the visual design space and potential of visual analysis, others exploring the same domain may arrive at a different set of conclusions. For example, in our study, we did not find strong evidence of the value of visualizing the input graph, despite precedence from prior work, and chose not to include it. Other researchers may find some value for the input graph in certain scenarios of analytic assets recommendations that did not appear in our study.  Our prioritization of recommendation probability and the visual design choices that support it, may not generalize to link-prediction tasks more broadly, which may have a different set of criteria that we do not capture.   
\vspace{.5em}
\section{Conclusion}
We conducted a design study that examined the validity of GNN-derived recommendations of analytic assets in the domain of data analysis. Through iterative collaboration with MLEs and SEs, we arrive at a set of design choices that we implement into the~\sys visual analysis tool. We furthermore contribute design choices and rationales that underlie~\sys and can generalize to broader applications for GNN-based recommendation systems. Our findings contribute to a growing understanding of how GNNs are used ``in the wild'' and the role that visual analysis plays in interrogating and evaluating the results of these models.

\bibliographystyle{abbrv-doi-hyperref}

\bibliography{main}

\end{document}